\documentclass[11pt,a4paper]{llncs}
\usepackage{fullpage,graphicx}
\newcommand{\BWT}{\ensuremath{\mathrm{BWT}}}
\newcommand{\occ}{\ensuremath{\mathrm{occ}}}
\newcommand{\LF}{\ensuremath{\mathrm{LF}}}

\begin{document}

\title{MIOV: Reordering MOVI for even better locality}

\author{Peter Pere\v{s}\'ini\inst{1} \and
	Nathaniel K.~Brown\inst{2} \and
	Travis Gagie\inst{3}
	\and Ben Langmead\inst{2}}

\institute{Comenius University, Bratislava, Slovakia \and
	Johns Hopkins University, Baltimore, USA \and
	Dalhousie University, Halifax, Canada}

\maketitle

M\"akinen, Navarro, Sir\'en and V\"alim\"aki~\cite{MNSV10} found that for highly repetitive datasets such as pangenomes, FM-indexes built atop run-length compressed Burrows-Wheeler Transforms (RLBWTs) can be much smaller than standard FM-indexes, while still supporting fast counting queries.  Policriti and Prezza~\cite{PP18} showed how to locate one match for a pattern $P$ quickly while still using $O (r)$ space, where $r$ is the number of runs in the BWT of the dataset, and Gagie, Navarro and Prezza~\cite{GNP20} showed how to locate all $\occ$ matches to $P$ quickly in the same space.  Nishimoto and Tabei~\cite{NT21} showed how to speed up key operations in Gagie et al.'s index by using table-lookup instead of rank queries on the RLBWT, which take predecessor-query time.  They thus achieved optimal $O (|P| + \occ)$ time when the size of the alphabet is polylogarithmic in the size of the dataset, still in $O (r)$ space.  There are now several implementations~\cite{BGR22,NKT22,ZBAGL24,BFN24} of Nishimoto and Tabei's index, which they called the {\em move structure}.  In particular, Zakeri et al.'s MOVI implementation keeps all the information about a run localized so that we can perform a backward-step, for example, not just in constant time time in theory but with only one cache miss in practice.

For each run $\BWT [i..k]$, MOVI stores a pointer to the run containing $\BWT [\LF (i)]$ and $\BWT [\LF (i)]$'s offset in that run, where $\LF (i)$ is the position in the BWT of the character that precedes $\BWT [i]$ in the dataset.  If we are backward-stepping from $\BWT [j]$ with $i \leq j \leq k$, then we look up the run containing $\BWT [\LF (i)]$ and $\BWT [\LF (i)]$'s offset $\ell$ in that run, then compute $\BWT [\LF (j)]$'s offset $\ell + j - i$ in that run.  If $\ell + j - i$ is greater than the length of the run then we subtract the length of the run from $\ell + j - i$ to get $\BWT [\LF (j)]$'s offset in the next run in the BWT, repeating this ``run-hopping'' as often as necessary.  (Nishimoto and Tabei showed how we can split the runs, without increasing the overall space by more than a constant factor, such that each backward-step requires a constant number of run-hops.)  If we also store pointers from each run to its successor in the BWT then we can view MOVI as an automaton with the runs as states --- and, for example, we can reorder the runs and still be able to invert the BWT and recover the dataset from it.  This is not a completely new idea: in implementations of Durbin's~\cite{Dur14} Positional BWT (PBWT), we improve locality by laying out in memory the compressed bitvectors for the columns in the same order the columns appear in the matrix; Sir\'en et al.~\cite{SGNPD20} use a similar idea in their Graph BWT (GBWT).  Indeed, Sir\'en first suggested this line of research to us.

\begin{figure}[t]
\begin{center}
\resizebox{.9\textwidth}{!}
{\begin{tabular}{c@{\hspace{5ex}}c@{\hspace{5ex}}c@{\hspace{8ex}}c@{\hspace{8ex}}c}
\begin{tabular}{cl}
BWT & context\\
\hline
\tt T  & \tt $\mathtt{\$}$AGATACA\\
\tt T  & \tt $\mathtt{\$}$GATACA\\
\tt T  & \tt $\mathtt{\$}$GATTACA\\
\tt T  & \tt $\mathtt{\$}$GATTAGA\\
\tt A  & \tt $\mathtt{\$}$GATTAGAT\\
\tt T  & \tt A$\mathtt{\$}$GATTAGA\\
\tt T  & \tt ACAT$\mathtt{\$}$AGA\\
\tt T  & \tt ACAT$\mathtt{\$}$GA\\
\tt T  & \tt ACAT$\mathtt{\$}$GAT\\
\tt T  & \tt AGAT$\mathtt{\$}$GAT\\
\tt T  & \tt AGATA$\mathtt{\$}$GAT\\
\tt $\mathtt{\$}$ & \tt AGATACAT\\
\tt C  & \tt AT$\mathtt{\$}$AGATA\\
\tt C  & \tt AT$\mathtt{\$}$GATA\\
\tt C  & \tt AT$\mathtt{\$}$GATTA
\end{tabular}
&
\begin{tabular}{cl}
BWT & context\\
\hline
\tt G  & \tt AT$\mathtt{\$}$GATTA\\
\tt G  & \tt ATA$\mathtt{\$}$GATTA\\
\tt G  & \tt ATACAT$\mathtt{\$}$A\\
\tt G  & \tt ATACAT$\mathtt{\$}$\\
\tt G  & \tt ATTACAT$\mathtt{\$}$\\
\tt G  & \tt ATTAGAT$\mathtt{\$}$\\
\tt G  & \tt ATTAGATA$\mathtt{\$}$\\
\tt A  & \tt CAT$\mathtt{\$}$AGAT\\
\tt A  & \tt CAT$\mathtt{\$}$GAT\\
\tt A  & \tt CAT$\mathtt{\$}$GATT\\
\tt A  & \tt GAT$\mathtt{\$}$GATT\\
\tt A  & \tt GATA$\mathtt{\$}$GATT\\
\tt A  & \tt GATACAT$\mathtt{\$}$\\
\tt $\mathtt{\$}$ & \tt GATACAT\\
\tt $\mathtt{\$}$ & \tt GATTACAT
\end{tabular}
&
\begin{tabular}{cl}
BWT & context\\
\hline
\tt $\mathtt{\$}$ & \tt GATTAGAT\\
\tt $\mathtt{\$}$ & \tt GATTAGATA\\
\tt A  & \tt T$\mathtt{\$}$AGATAC\\
\tt A  & \tt T$\mathtt{\$}$GATAC\\
\tt A  & \tt T$\mathtt{\$}$GATTAC\\
\tt A  & \tt T$\mathtt{\$}$GATTAG\\
\tt A  & \tt TA$\mathtt{\$}$GATTAG\\
\tt A  & \tt TACAT$\mathtt{\$}$AG\\
\tt A  & \tt TACAT$\mathtt{\$}$G\\
\tt T  & \tt TACAT$\mathtt{\$}$GA\\
\tt T  & \tt TAGAT$\mathtt{\$}$GA\\
\tt T  & \tt TAGATA$\mathtt{\$}$GA\\
\tt A  & \tt TTACAT$\mathtt{\$}$G\\
\tt A  & \tt TTAGAT$\mathtt{\$}$G\\
\tt A  & \tt TTAGATA$\mathtt{\$}$G
\end{tabular}
&
\begin{tabular}{r|ccrc}
& chr & len & ptr & off \\
\hline\\[-2ex]
 1) & \tt T  & 4 &  9 & 1 \\
 2) & \tt A  & 1 &  3 & 1 \\
 3) & \tt T  & 6 &  9 & 5 \\
 4) & \tt \$ & 1 &  1 & 1 \\
 5) & \tt C  & 3 &  7 & 1 \\
 6) & \tt G  & 7 &  7 & 4 \\
 7) & \tt A  & 6 &  3 & 2 \\
 8) & \tt \$ & 4 &  1 & 2 \\
 9) & \tt A  & 7 &  5 & 1 \\
10) & \tt T  & 3 & 11 & 1 \\
11) & \tt A  & 3 &  6 & 5
\end{tabular}
&
\begin{tabular}{r|ccrcr}
& chr & len & ptr & off & succ\\
\hline\\[-2ex]
 1) & \tt T  & 4 &  4 & 1 &  2\\
 2) & \tt A  & 1 &  3 & 1 &  3\\
 3) & \tt T  & 6 &  4 & 5 &  9\\
 4) & \tt A  & 7 &  5 & 1 & 10\\
 5) & \tt C  & 3 &  7 & 1 &  6\\
 6) & \tt G  & 7 &  7 & 4 &  7\\
 7) & \tt A  & 6 &  3 & 2 &  8\\
 8) & \tt \$ & 4 &  1 & 2 &  4\\
 9) & \tt \$ & 1 &  1 & 1 &  5\\
10) & \tt T  & 3 & 11 & 1 & 11\\
11) & \tt A  & 3 &  6 & 5 &  1
\end{tabular}
\end{tabular}}
\caption{A simplified BWT {\bf (left)}, move structure {\bf (center)} for {\tt GATTACAT\$}, {\tt AGATACAT\$}, {\tt GATACAT\$}, {\tt GATTAGAT\$}, {\tt GATTAGATA\$} and reordered move structure {\bf (right)}.}
\label{fig:BWT}

\medskip

\includegraphics[width=.4\textwidth]{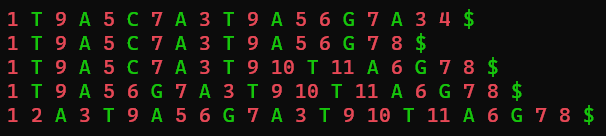}
\hspace{5ex}
\includegraphics[width=.4\textwidth]{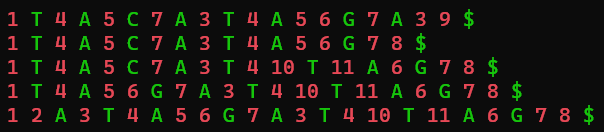}
\end{center}
\caption{The runs we visit while inverting the BWT from Figure~\ref{fig:BWT} before reordering {\bf (left)} and after {\bf (right)}.  As usually with a BWT, the strings are recovered reversed.}
\label{fig:outputs}
\end{figure}

Good reorderings for the PBWT and GBWT are fairly obvious, but it is not clear how we should reorder MOVI.  For small datasets like the one shown in Figure~\ref{fig:BWT}, we can find the optimal reordering exactly with binary integer linear programming.  To see why, consider how we visit the runs while inverting the BWT, shown on the left in Figure~\ref{fig:outputs}.  We move from run 1 to run 2 once, from run 1 to run 9 four times, from run 2 to run 3 once, from run 3 to run 9 six times, etc..  Overall, we move from one run to the next run memory $23 / 53 \approx 43\%$ of the time.  If we reorder the runs such that the current run 3 appears immediately before the current run 9, then the six moves will be more local.  Therefore, if we say $x_{i, j}$ is a binary indicator variable reflecting whether run $i$ precedes run $j$ in  memory, we can find the optimal reordering by maximizing
\begin{eqnarray*}
\lefteqn{x_{1, 2} + 4\,x_{1, 9} + x_{2, 3} + x_{3, 4} +
6\,x_{3, 9} + 4\,x_{5, 6} + 3\,x_{5, 7} + \mbox{}}\\
&& 7\,x_{6, 7} + 6\,x_{7, 3} + 4\,x_{7, 8} + 7\,x_{9, 5} +
3\,x_{9, 10} + 3\,x_{10, 11} + 3\,x_{11, 6}
\end{eqnarray*}
subject to constraints requiring that the $x_{i, j}$s reflect a total order, 
\[\sum_i x_{i, j} \leq 1\,, \hspace{2ex}
\sum_j x_{i, j} \leq 1\,, \hspace{2ex}
y_{i, j} + y_{j, i} = 1\,, \hspace{2ex}
y_{i, j} + y_{j, k} + y_{k, i} \leq 2\,, \hspace{2ex}
x_{i, j} + y_{j, i} \leq 1\,.\]
One optimal solution is to swap runs 4 and 9, after which we visit the runs as shown on the right in Figure~\ref{fig:outputs}, moving from one run to the next run in memory $33 / 53 \approx 62\%$ of the time --- an increase of 20\%.

Binary integer linear programming is unlikely to scale to datasets whose BWTs have billions of runs, of course, so we are now looking at other ways to find good reorderings.  For example, we are now experimenting with storing information for the runs in the order they are first visited when inverting the BWT.  Our preliminiary experimental results are interesting, and we hope to have publishable results in the near future.

\end{document}